\documentclass[pre,twocolumn]{revtex4}
\usepackage{amsmath}
\usepackage{epsfig}

\begin{document}

\title{Dielectric response of a polar fluid trapped in a spherical nanocavity}

\author{Ronald Blaak}
\email{rb419@cam.ac.uk}
\affiliation{Department of Chemistry, University of Cambridge,
  Lensfield Road, Cambridge, CB2 1EW, United Kingdom}  

\author{Jean-Pierre Hansen}
\email{jph32@cam.ac.uk}
\affiliation{Department of Chemistry, University of Cambridge,
 Lensfield Road, Cambridge, CB2 1EW, United Kingdom} 

\begin{abstract}
We present extensive Molecular Dynamics simulation results for the
structure, static and dynamical response of a droplet of 1000 soft
spheres carrying extended dipoles and confined to spherical cavities
of radii $R=2.5$, 3, and 4 nm embedded in a dielectric continuum of
permittivity $\epsilon' \geq 1$. The polarisation of the external
medium by the charge distribution inside the cavity is accounted for
by appropriate image charges. We focus on the influence of the
external permittivity $\epsilon'$ on the static and dynamic properties
of the confined fluid. The density profile and local orientational
order parameter of the dipoles turn out to be remarkably insensitive
to $\epsilon'$. Permittivity profiles $\epsilon(r)$ inside the
spherical cavity are calculated from a  generalised Kirkwood
formula. These profiles oscillate in phase with the density profiles
and go to a ``bulk'' value $\epsilon_b$ away from the confining
surface; $\epsilon_b$ is only weakly dependent on $\epsilon'$, except
for $\epsilon' = 1$ (vacuum), and is strongly reduced compared to the
permittivity of a uniform (bulk) fluid under comparable thermodynamic
conditions. 

The dynamic relaxation of the total dipole moment of the sample is
found to be strongly dependent on $\epsilon'$, and to exhibit
oscillatory behaviour when $\epsilon'=1$; the relaxation is an order
of magnitude faster than in the bulk. The complex frequency-dependent
permittivity $\epsilon(\omega)$ is sensitive to $\epsilon'$ at low
frequencies, and the zero frequency limit  $\epsilon(\omega=0)$ is
systematically lower than the ``bulk'' value $\epsilon_b$ of the
static primitivity.
\end{abstract}

\maketitle

\section{Introduction}
Following the pioneering work of Debye~\cite{Book:Debye},
Kirkwood~\cite{Kirkwood:1939JCP}, and Onsager~\cite{Onsager:1936JACS}, 
the bulk dielectric response of polar materials is by now well
understood~\cite{Madden:1984ACP}, and dielectric response is a method
of choice for the experimental investigation of molecular dynamics in
condensed matter. Simulations of model polar systems have played a key
role in our understanding of dielectric fluids~\cite{Weis:2005APS},
and the subtle problems arising in the simulation of finite, but
periodically repeated samples of such fluids, linked to the proper
handling of boundary conditions, have been clarified in the
eighties~\cite{Neumann:1983MP,Neumann:1983CPL}. However, with few
exceptions~\cite{Bagchi:1991ACP,Senapati:1999JCP,Teschke:2001PRE,Stern:2003JCP,Ballenegger:2003EPL,Ballenegger:2005JCP},
much less experimental, theoretical and numerical effort has gone into
understanding the dielectric response of confined polar fluids.

Consider a fluid of polar molecules trapped in a finite or infinite
(along one or two directions) cavity surrounded by a dielectric
material characterised by a permittivity different from that of the bulk
polar fluid. The global dielectric response of the former is determined
by the fluctuations and relaxation of the overall dipole moment of the
trapped fluid. The question we wish to address is how the dipolar fluctuations
are affected by confinement, i.e. by the presence of a surface
separating the polar fluid from the dielectric material which
surrounds the cavity. One can distinguish between two main effects due
to the presence of such interfaces. The first is purely geometric:
how are the dipolar fluctuations affected in the vicinity of a
non-polarisable interface, compared to bulk fluctuations? In
particular, can one define a meaningful local dielectric permittivity
tensor $\tensor{\epsilon}(\vec{r})$, when the polar molecules are
restricted to stay on one side of a confining surface, with vacuum on
the other side. This question has been recently addressed in the case
of simple geometries (slab or spherical
cavity)~\cite{Senapati:1999JCP,Ballenegger:2005JCP}. The second effect
arises from the electric boundary conditions which must be satisfied when
the confining medium is polarisable, and hence characterised by a
permittivity $\epsilon'>1$. In this paper we investigate the second
effect in the case of a simple polar fluid confined to a spherical cavity
carved out of a dielectric continuum which extends to infinity in all
directions. Note that in the limit where the system consists of a
single polar molecule fixed at the centre of the spherical cavity, the
system reduces to Onsager's celebrated model for the calculation of
the permittivity of a polar material~\cite{Onsager:1936JACS}.

The model of a polar fluid in a spherical cavity may be regarded as a
crude representation of dual physical situations. One concerns
micro-emulsions where inverse micelles are nanodroplets of water in
oil, which are stabilised by a monolayer of surfactants. The majority
oil phase then provides the embedding dielectric medium. The conjugate
situation is that of a globular macromolecule (i.e. a protein), made
up of polar segments, dissolved in water. The connectivity of the
macromolecule is then crudely accounted for by confining the
unconnected polar residues to a spherical volume equal to that of the
cavity. In that case the solvent (water) plays the role of the
embedding continuum.

\section{Model and simulation methodology}
\label{S:model}
Consider a system of $N$ polar molecules confined to a spherical cavity
of radius $R$ surrounded by an infinite dielectric continuum of
permittivity $\epsilon'$. Following related earlier
work~\cite{Ballenegger:2004MP,Ballenegger:2005JCP} the model which
will be investigated is one of spherical molecules carrying
extended (rather than point) dipoles consisting of two opposite
charges $\pm q$ displaced symmetrically by a distance $d/2$ from the
centre of the molecule, such that the absolute dipole moment is $\mu =
q d$. Let $\vec{r}_i$ be the position of the centre of the molecule $i$,
and $\hat{\mu}_i$ be the unit vector along the dipole moment of that
molecule. The two charges $q_\pm = \pm q$ are than placed at
$\vec{r}_{i\pm} = \vec{r}_i \pm \frac{d}{2} \hat{\mu}_i$.

If $\phi(\vec{r},\vec{r}')$ is the electrostatic potential at
$\vec{r}'$ due to a unit charge at $\vec{r}$, taking proper account
of the electrostatic boundary conditions at the surface of the
spherical cavity, then the total interaction energy of a pair of
molecules $i$ and $j$ is: 
\begin{equation}
v(\vec{r}_i,\vec{r}_j) = v_0(|\vec{r}_i-\vec{r}_j|) +
\sum_{\alpha,\beta =\pm} q_\alpha q_\beta
\phi(\vec{r}_{i\alpha},\vec{r}_{j\beta}) 
\end{equation}
where $v_0(r)$ is the short-range repulsive potential between the
spherical molecules, which is chosen to be of inverse power form as
in~\cite{Ballenegger:2005JCP}:
\begin{equation}
\label{E:v0}
v_0(r) = 4 u \left( \frac{\sigma}{r} \right)^n
\end{equation}
with $n=12$ in practice.

The exact form of $\phi(\vec{r},\vec{r}')$ for the spherical geometry
is derived in the Appendix by solving Poisson's equation with the
appropriate electrostatic boundary conditions. This cumbersome
expression is not well adapted to simulations, and may be replaced by
the approximation:
\begin{equation}
\label{E:phip}
\phi(\vec{r},\vec{r}') = \frac{1}{4 \pi \epsilon}
\left[\frac{1}{|\vec{r}-\vec{r}'|} + (1 - 2 \kappa)
  \frac{(R/r)}{|\vec{r}^* -\vec{r}'|} \right] 
\end{equation}
where $\vec{r}^* = (R/r)^2 \vec{r}$, $\epsilon$ is the permittivity of
the empty cavity ($\epsilon=\epsilon_0$ in practice) and
$\kappa=\epsilon'/(\epsilon+\epsilon')$. The potential $\phi$ is seen
to reduce to the bare Coulomb potential when $\epsilon=\epsilon'$, 
i.e. in the absence of a dielectric discontinuity, and reduces to the
classic result for a cavity surrounded by a conductor (metallic
boundary condition $\epsilon ' = \infty$)~\cite{Book:Jackson}. In the
approximation~(\ref{E:phip}) the image charge is located at the same
position as in the metallic boundary case, but its weight differs from
-1.

Atoms outside the sphere, making up the dielectric continuum of
permittivity $\epsilon'$, are assumed to interact with the dipolar molecules
inside the cavity by the short-range potential $v_0(r)$ in
Eq.~(\ref{E:v0}). These atoms are assumed to be distributed
uniformly with a number density $\rho$, so that the external potential
acting on the molecules within the cavity is:
\begin{equation}
V_{ext}(r) = \rho \int_R^\infty  d r' r'^2 \int_0^\pi d \theta \sin
\theta \int_0^{2\pi} d \phi ~ 4 u \left( \frac{\sigma}{\bar{r}}
\right)^n 
\end{equation}
where $\bar{r}^2 = r^2 + r'^2 - 2 r r' \cos \theta$. A straightforward
integration leads to:
\begin{equation}
\begin{split}
V_\text{ext}(r)= & \frac{8\pi u \rho \sigma^n}{(n-2)(n-3)(n-4)} \frac{1}{r}
  \times \\ 
 & \\ 
  & \left[ \frac{(n-3)R-r}{(R-r)^{n-3}} - \frac{(n-3)R+r}{(R+r)^{n-3}}
  \right] 
\end{split}
\end{equation}
The external potential goes through a minimum at the origin, so that
the external force vanishes at $r=0$, as expected by symmetry. In
practice the reduced density $\rho^*=\rho \sigma^3$ of the dielectric
continuum is chosen to be equal to 1, thus mimicking a dense medium.

The coupled classical equations of motion for the translations of the
molecular centres and the rotations of the dipoles were solved by a
standard velocity Verlet algorithm, using the GROMACS Molecular
Dynamics (MD) package\cite{Gromacs} under constant temperature
conditions, imposed by a Berendsen thermostat and with a time-step $\Delta
t= 1$ fs. The values of the key physical parameters are listed in
Table~\ref{T:par}. Most simulations were carried out for samples of
$N=1000$ molecules and for three cavity radii $R=4$, 3, and 2.5
nm. Nominal overall densities may be estimated as $\rho_0=3N/(4 \pi
R_\text{eff}^3)$ where $R_\text{eff}<R$ is an effective radius of the
cavity. The latter may be estimated from the radial density profiles
$\rho(r)$, to be introduced in the following section, by requiring
$\rho(r=2 R_\text{eff} - R) = \rho(r=0)$. This makes the effective
radius of the cavity roughly half a particle diameter smaller than the
radius at which the dielectric medium starts. It is convenient to
introduce the following reduced variables:  
\begin{table}[thb]
\begin{tabular}{l|l|ll}
parameter & & \\
\hline
time-step         & $\Delta t$ & 1          & fs    \\
dipole charge     & $q$        & 0.41843035 & e     \\
charge separation & $d$        & 0.12190214 & nm    \\
diameter          & $\sigma$   & 0.36570642 & nm    \\
energy scale      & $u$        & 1.8476692  & kJ/mol\\
temperature       & $T$        & 300        & K     \\
dipole            & $\mu$      & 2.4500000  & Debye \\
mass              & $m$        & 10         & a.m.u.
\end{tabular}
\caption{Physical parameters as used in the simulations. Both charges
  $\pm q$ of the molecule carry a mass of 5 a.m.u.}
\label{T:par}
\end{table}
\begin{table}[thb]
\begin{tabular}{l|l}
Reduced units &  \\
\hline
 $d^*$    & 0.3333    \\
 $\mu^*$  & 2         \\
 $I^*$    & 0.0278    \\
 $T^*$    & 1.35      \\
\end{tabular}
\caption{Key parameters in reduced units}
\label{T:units}
\end{table}
\begin{equation}
\begin{split}
\mu^*    & = \sqrt{\frac{\mu^2}{(4 \pi \epsilon_0) u \sigma^3}} \\
d^*      & = \frac{d}{\sigma}            \\
T^*      & = \frac{k_B T}{u}             \\
\rho_0^* & = \rho_0 \sigma^3             \\
I^*      & = \frac{I}{m\sigma^2} = \frac{1}{4}d^{*2}
\end{split}
\end{equation}
Values of these reduced variables used in the simulations are listed
in Table~\ref{T:units}. Runs extended over several million time-steps,
corresponding to phase space trajectories of several ns.

The model considered in this paper is not unlike that investigated by
Senapati and Chandra~\cite{Senapati:1999JCP}, who used the Stockmayer
potential for much smaller systems ($N \simeq 100$), and restricted
their calculations to the case $\kappa=0.5$, i.e. to a cavity
surrounded by vacuum.

\section{Static properties}
\label{S:static}

This section focuses on the results of our MD simulations for the
structure and static dielectric response of the model defined in
Sect.~\ref{S:model} for a polar fluid in a spherical cavity. We have
considered embedding dielectric continua of 
permittivities $\epsilon'=1$ (vacuum), 4, 9, and $\infty$ (metal)
corresponding to values of the 
parameter $\kappa=\epsilon'/(\epsilon+\epsilon')=0.5$, 0.8, 0.9, and
1. The structure of the polar fluid is conveniently characterised by
the radial density profile $\rho(r)$, where $r$ is the distance of the
centre of a polar molecule from the centre of the cavity; clearly
$\rho(r)=0$ for $r>R$. The computed profiles integrate up to the total
number of polar molecules in the cavity:
\begin{equation}
4 \pi \int_0^R d r \rho(r) r^2  = N
\end{equation}
\begin{figure}[htb]
\centering
\includegraphics[width=8cm]{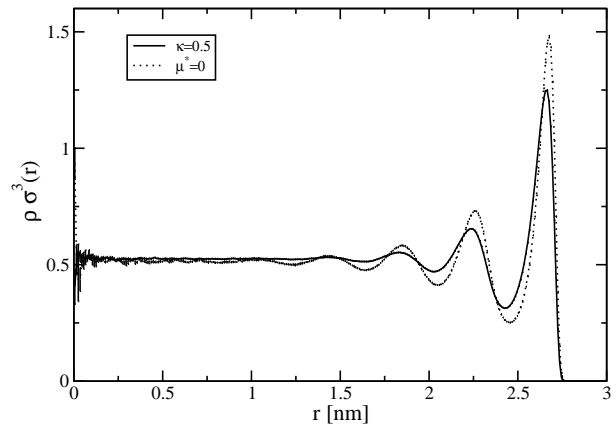}
\caption{Density profiles $\rho(r)$ as functions of the radial
  distance $r$ for a system with $N=1000$ and $R=3$ nm
  ($\rho^*=0.53$). The solid curve is for dipoles ($\mu^*=2$) with
  $\kappa=0.5$. The different values of $\kappa$ coincide on
  this scale. The dotted curve is the reference for
  particles without dipoles ($\mu^*=0$).} 
\label{F:rho}
\end{figure}
Profiles obtained for a cavity of radius $R=3$ nm, $N=1000$, $\mu^*=2$
and four values of $\kappa$ are compared in Fig.~\ref{F:rho} to the
profile corresponding to non-polar molecules ($\mu^*=0$) under
otherwise identical conditions. All profiles exhibit the expected
layering near the confining spherical surface~\cite{Senapati:1999JCP}. As expected the
layering effect is even more pronounced for the smaller cavity radius
$R=2.5$ nm which we also explored (data not shown). There are two
striking results: the profiles observed for the four different values
of $\kappa$ are nearly indistinguishable, i.e. the radial structure turns
out to be practically independent of the polarisability of the
confining continuum. However the profile $\rho(r)$ changes
substantially when $\mu^*$ is set equal to zero, i.e. in the absence
of any electrostatic coupling between molecules and with the embedding
medium: the layering is seen in Fig.~\ref{F:rho} to be considerably
enhanced, and to extend deeper towards the centre of the cavity. 
\begin{figure}[thb]
\centering
\includegraphics[width=8cm]{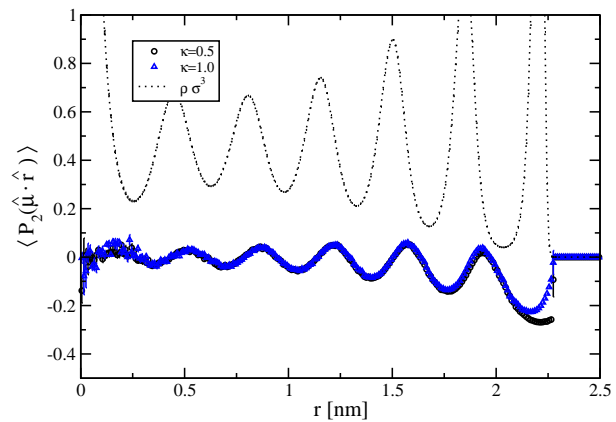}
\caption{The order parameter $\langle P_2(\hat{\mu} \cdot \hat{r})
  \rangle$ as a function of the distance to the centre of the cavity
  ($N=1000$, $R=2.5$ nm). The dotted curve shows the corresponding
  density profile $\rho^*(r)$ for $\kappa=0.5$.}
\label{F:P2mun}
\end{figure}
\begin{figure}[thb]
\centering
\includegraphics[width=8cm]{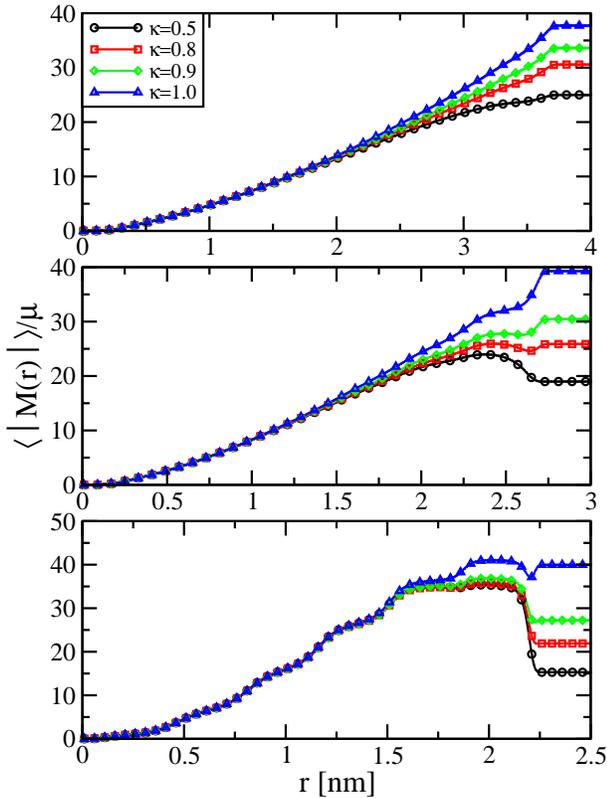}
\caption{The average absolute value $\langle |\vec{M}(r)| \rangle$ of
  the total dipole moment of all dipoles within a distance $r$ from
  the centre of the cavity, for the various values of $\kappa$. From
  top to bottom cavities of size $R=4$, 3, and 2.5 nm.}
\label{F:M}
\end{figure}
These
findings are qualitatively confirmed in the cases of the larger ($R=4$
nm) and smaller ($R=2.5$ nm) cavities. The conclusion to be drawn here
is that the dipolar interactions between molecules tend to smooth out
the layering imposed by the short-range, excluded volume effects. 
The orientations of individual dipoles relative to the normal to the
surface are characterised by the local order parameters
$\langle P_1(\hat{\mu} \cdot \hat{r}) \rangle_r$ and $\langle
P_2(\hat{\mu} \cdot \hat{r}) \rangle_r$, where $P_l$ denotes the $l$th
order Legendre polynomial, $\hat{r}$ and $\hat{u}$ are the unit
vectors along the radial vector $\vec{r}$ and the dipole moment
$\vec{\mu}$, and the statistical average is taken over dipole
configurations within spherical shells of radius $r$ and width
$\delta$. Because of the $\vec{\mu}_i \rightarrow -\vec{\mu}_i$
inversion symmetry, $\langle P_1(\hat{\mu} \cdot \hat{r}) \rangle_r$
is identically zero, while $\langle P_2(\hat{\mu} \cdot \hat{r})
\rangle_r$ is plotted in Fig.~\ref{F:P2mun} as a function of $r$ for
several values of $\kappa$. $\langle P_2(\hat{\mu} \cdot \hat{r})
\rangle_r$ is seen to depend very little on $\kappa$, except very
close to the outer surface. The order parameter oscillates
somewhat out of phase with the oscillations in the density profile
shown in a frame of the same figure. Near the maxima of the latter,
which correspond to well defined shells of polar molecules, the order
parameter is predominantly negative, signalling a preferential
orientation of the dipoles orthogonal to the radial vector $\hat{r}$,
i.e. the dipoles orient preferentially parallel to the confining
surface, irrespective of the embedding medium, suggesting a vortex-like
pattern of the confined dipoles.

Qualitatively similar behaviour is observed for the larger cavities
(lower overall densities), except that, as expected, the oscillations
in $\langle P_2(\hat{\mu} \cdot \hat{r}) \rangle_r$ are less
pronounced. The preferential alignment of dipoles parallel to the
confining surface was also observed in earlier work~\cite{Zhang:1995JPC}.

Consider next the total dipole moment $\vec{M}(r)$ within a sphere of
radius $r\leq R$. While $\langle \vec{M}(r) \rangle$ vanishes again by
symmetry, the statistical average of the absolute value of
$\vec{M}(r)$ shows an interesting behaviour, illustrated in
Fig.~\ref{F:M} for the three different pore radii $R$ under
investigation. For the two lower densities  $\langle |\vec{M}(r)
| \rangle$ is seen to increase roughly as $N_r^{1/2}$ i.e. $r^{3/2}$,
up to $r \simeq R/2$, as one would expect if the $N_r$ dipole moments
within a sphere of radius $r$ were uncorrelated. 
This part of the curve is essentially independent of $\kappa$. 
\begin{figure}[thb]
\centering
\includegraphics[width=8cm]{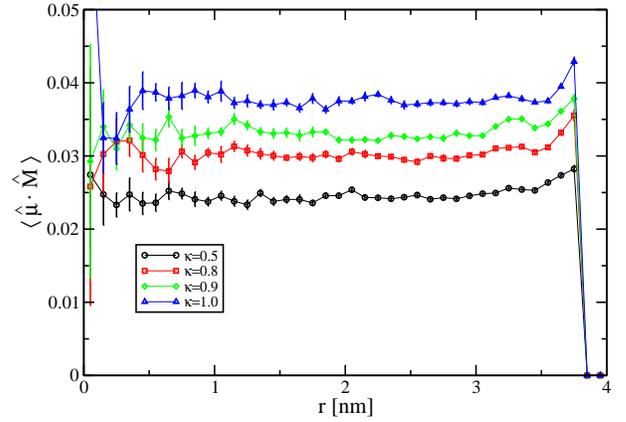}
\caption{The correlation $\langle \hat{\mu} \cdot \hat{M} \rangle$ of
  the individual dipoles with the direction of the total dipole moment
  of all particles in the cavity as a function of the distance $r$ from
  the centre for a $R=4$ nm cavity, and $0.5 \leq \kappa\leq1$.}
\label{F:P1muM}
\end{figure}
Beyond 
$r \simeq R/2$ the four curves diverge and show some structure for the
smaller cavities ($R=3$ and 2.5 nm). For a given $r$, the value of
$\langle |\vec{M}(r)| \rangle$ increases with increasing $\kappa$,
reflecting an enhanced effect of the image dipoles as $\epsilon'$
increases (See. Eq.~\ref{E:phip}). Although not evident in the density
profiles of Fig.~\ref{F:rho}, the effect of the polarisation of the
confining medium is seen to have a very significant effect on the
dipolar properties of the confined system. This is also evident in
Fig.~\ref{F:P1muM} where we plot the average of the projection of the
dipole moments of the individual dipoles within a shell of radius $r$
and small thickness
along the total dipole moment $\vec{M}$ of the 
sample, $\langle \hat{\mu}\cdot \hat{M}\rangle$. The projection is
small, but positive signalling that the individual dipoles tend to
align along the overall dipole moment.
Moreover, it is quasi-independent of the distance
$r$, and increases with $\kappa$, i.e. as one moves from vacuum
outside the cavity, to a metallic confining medium.

We finally turn to the static dielectric permittivity profile of the
confined fluid. This can be related to the dipolar fluctuations by
linear response theory~\cite{Ballenegger:2005JCP}, or by measuring the
total polarisation induced in the sample by the ``external'' electric
field due to a charge placed at the origin of the cavity. Let
$\vec{m}(r)$ denote the microscopic polarisation density:
\begin{equation}
\vec{m}(r) = \sum_{i=1}^N \vec{\mu}_i \delta(\vec{r}-\vec{r}_i)
\end{equation}
The overall dipole moment of the sample is then:
\begin{equation}
\vec{M} = \int_{{\cal D}_\text{cavity}} d \vec{r} \vec{m}(r)
\end{equation}
where the integration is over the whole volume of the cavity. The
linear response result for the permittivity profile $\epsilon(r)$ is
given by the following generalisation~\cite{Ballenegger:2005JCP} of
Kirkwood's classical results for the
bulk~\cite{Kirkwood:1939JCP}:
\begin{equation}
\label{E:ep}
\frac{(\epsilon(r)-1)(2 \epsilon' + 1)}{2 \epsilon' + \epsilon(r)} = 
\frac{4 \pi \beta}{3} \left[ \langle \vec{m}(r) \cdot \vec{M} \rangle
  - \langle \vec{m}(r) \rangle \cdot \langle \vec{M}\rangle \right]
\end{equation}
Far from the confining surface bulk behaviour may be expected, and
replacing $\vec{m}$ by $\vec{M}/V$, one recovers Kirkwood's
formula. Near the surface rotational invariance is broken and the
permittivity becomes a tensor with longitudinal (i.e. parallel to
$\vec{r}$) and transverse components.

Note that equation (\ref{E:ep}) is exact provided that there exists a
{\em local} relationship between the polarisation and the internal
(Maxwell) electric field. It defines a permittivity profile
$\epsilon(r)$ which may be expected to go to a constant ``bulk'' value
far from the confining surface, as will be confirmed by our simulation
data, at least at low or moderate densities. An approximate method for
estimating such ``bulk'' values within cavities has been put forward
by Berendsen~\cite{Berendsen,Adams:1976MP}, but the limitations of
this method have been illustrated in Ref.\cite{Ballenegger:2005JCP}.

\begin{figure}[thb]
\centering
\includegraphics[width=8cm]{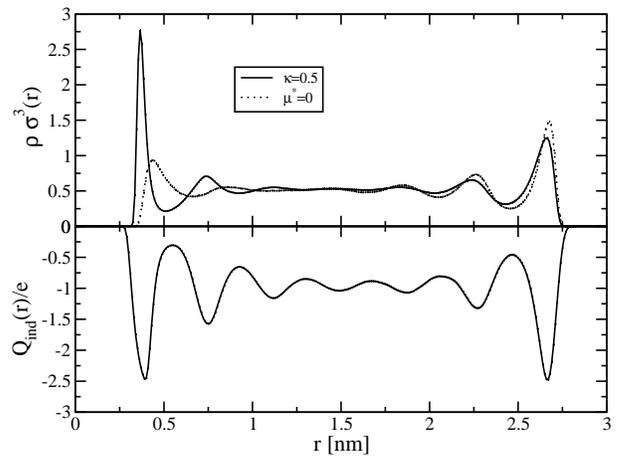}
\caption{The radial density profiles $\rho(r)$ in the presence of a
  charge at the centre of the cavity for $\kappa=0.5$ ($R=3$ nm,
  $\mu^*=2$).  The different values of $kappa$ coincide on
  this scale. The dotted curve is the reference
  profile for particles without dipole. In the lower part of the
  figure the corresponding charge distribution $Q_\text{ind}(r)/e$ is shown.}
\label{F:rhoQ30}
\end{figure}

In MD simulations, the correlation function on the right hand side of
the Eq.~(\ref{E:ep}) is estimated by averaging over all dipole moments
of particles within a spherical shell of radius $r$ and 
width $\simeq \sigma$.
In the ``external
field'' method, an additional particle is 
placed at the origin, with the (extended) dipole replaced by a simple
proton charge $e$ at its centre. If one assumes a local relationship
between the radial polarisation density $P(r) = \langle \hat{r} \cdot
\vec{m}(\vec{r}) \rangle$ and the local radial electric field
$E(r) = \hat{r} \cdot \vec{E}(\vec{r})$ of the form:
\begin{equation}
\label{E:P}
P(r)=\epsilon_0 \chi(r) E(r) = [ \epsilon(r)-1]  E(r)
\end{equation}
then, $\epsilon(r)$ follows from elementary
electrostatics~\cite{Ballenegger:2005JCP}. Let $Q(r)=e +
Q_\text{ind}(r)$ be the total charge contained inside a sphere of
radius $r$, which is easily estimated from the MD simulations for the
extended dipole model. As a consequence of the divergence theorem,
$Q_\text{ind}(r)$ inside the sphere of radius $r$ is related to the
polarisation density by:
\begin{equation}
\label{E:Pr}
P(r)=-\frac{Q_\text{ind}(r)}{4 \pi r^2}
\end{equation}
while the electric field $E(r)$ is related to $Q(r)$ by:
\begin{equation}
\label{E:Er}
E(r)= \frac{Q(r)}{r^2}
\end{equation}
Substitution of (\ref{E:Pr}) and (\ref{E:Er}) in Eq.~(\ref{E:P}) leads
to the desired estimate:
\begin{equation}
\label{E:epr}
\epsilon(r)= \frac{e}{e - 4 \pi r^2 P(r)} = \frac{1}{1 +
  Q_\text{ind}(r)/e} 
\end{equation}
The presence of the central charge introduces some distortion of the
density profiles near the centre of the cavity, as illustrated in
Fig.~\ref{F:rhoQ30}. An
excluded volume zone and subsequent layering now appear at small
$r$, but the profiles are virtually unchanged for $r \gtrsim R/2$
relative to the case without central charge. Note that adding the
additional particle changes the overall density inside the cavity by
only one part in 1000, so that a comparison between fluctuation and
response results remains meaningful.

\begin{figure}[thb]
\centering
\includegraphics[width=8cm]{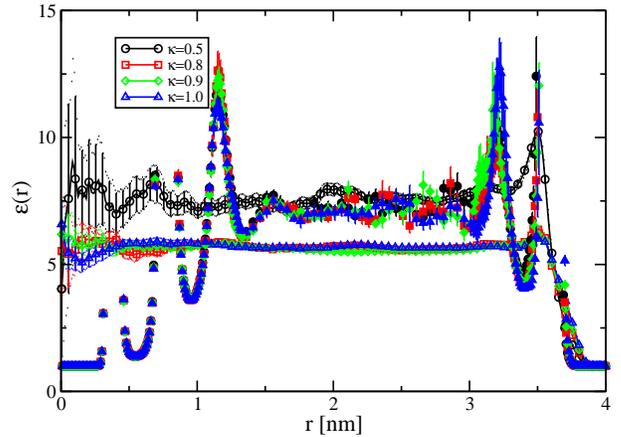}
\caption{$\epsilon(r)$ versus the distance $r$ obtained from the
  fluctuation formula (\ref{E:ep}) (open symbols) and from the response
  to a central charge (\ref{E:epr}) (filled symbols) for $\mu^*=2$, $R=4$nm,
  and $0.5 \leq \kappa \leq 1$.} 
\label{F:ep40}
\end{figure}

Plots of the induced charge $Q_\text{ind}(r)$ inside a sphere of
radius $r$ are shown as a function of $r$ in Fig.~\ref{F:rhoQ30} for the
cavity with radius $R=3$ nm. $Q_\text{ind}(r)$
``overscreens'' the external charge $e$ at the centre, at short
distances, before oscillating around a negative value and going to
zero as $r \rightarrow R$ . At the lower overall density ($R=4$ nm),
$Q_\text{ind}(r)$ stabilises around a ``bulk'' value above $-e$ at intermediate
distances (data not shown), while no such ``bulk'' regime is observed at higher density
($R=3$ nm) . Remarkably $Q(r)$ is practically independent of
$\epsilon'$ (or $\kappa$).

\begin{figure}[thb]
\centering
\includegraphics[width=8cm]{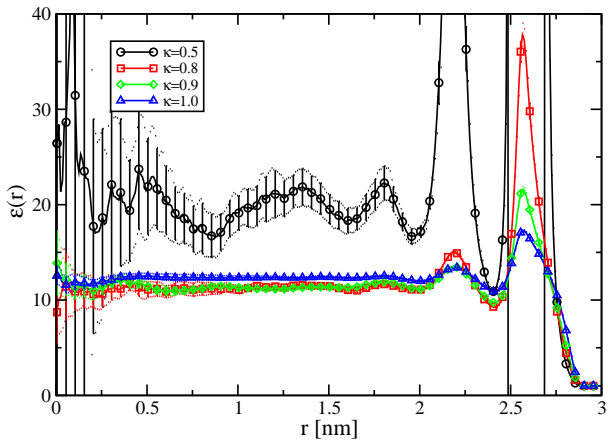}
\caption{$\epsilon(r)$ versus the distance $r$ obtained from the
  fluctuation formula (\ref{E:ep}) for $\mu^*=2$, $R=3$ nm, and $0.5
  \leq \kappa \leq 1$.} 
\label{F:ep30}
\end{figure}
\begin{figure}[thb]
\centering
\includegraphics[width=8cm]{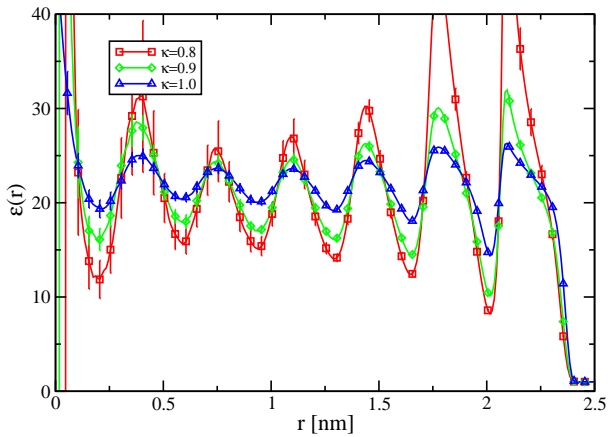}
\caption{$\epsilon(r)$ versus the distance $r$ obtained from the
  fluctuation formula (\ref{E:ep})for $\mu^*=2$, $R=2.5$ nm, and
  $\kappa=0.8$, 0.9, and 1.}
\label{F:ep25}
\end{figure}

The values of $\epsilon(r)$ derived from Eq.~(\ref{E:epr}) are
compared to the corresponding values estimated from the fluctuation
formula (\ref{E:ep}) in Fig.~\ref{F:ep40} for the lower density ($R=4$
nm), $\mu^*=2$ and four values of $\kappa$. Clearly Eq.~(\ref{E:epr})
can only yield physically acceptable results as long as
$Q_\text{ind}(r)/e > -1$. At small and large $r$ the oscillations in
$Q(r)$ lead to unphysical values of $\epsilon(r)$, signalling the
break-down of the local assumption (\ref{E:P}), as already noted in
Ref.~\cite{Ballenegger:2005JCP} for the special case $\kappa=0.5$
($\epsilon'=1$). In that case the data shown in Fig.~\ref{F:ep40}
point, however, to two surprising findings. First of all
$\epsilon(r)$ profiles derived from Eq.~(\ref{E:epr}) appear to be
independent of $\kappa$, due to the quasi-independence of
$Q_\text{ind}(r)$ on $\kappa$ in Fig.~\ref{F:rhoQ30}. On the other hand the
$\epsilon(r)$ values derived from the fluctuation formula (\ref{E:ep})
for $\epsilon'>1$ ($\kappa=0.8$, 0.9, and 1) are practically
independent of $\kappa$, but lie significantly below the results for
$\kappa=0.5$.  In other words, the polarisation of the surrounding
medium leads to a reduction of the permittivity of the sample inside
the cavity. However the response to a central charge $e$ appears to be
significantly non-linear, but independent of $\kappa$. It would be
very difficult to measure the response to a smaller external charge at
the cavity centre, due to an unfavourable signal to noise ratio.

\begin{table}[htb]
\begin{tabular}{l|l|l|l|l}
$R$     & $\rho^*$ & $\kappa$ & $\epsilon_b$ & $\epsilon(\omega=0)$ \\
\hline
4 nm    & 0.23  & 0.5 & 7.6  &  4.8 \\
        &       & 0.8 & 5.7  &  4.5 \\
        &       & 0.9 & 5.6  &  4.6 \\
        &       & 1.0 & 5.7  &  4.8 \\
3 nm    & 0.53  & 0.5 & 19.7 & 10.5 \\
        &       & 0.8 & 11.3 &  9.0 \\
        &       & 0.9 & 11.3 &  9.5 \\
        &       & 1.0 & 12.3 & 10.7 \\
2.5 nm  & 0.92  & 0.5 &      & 18.2 \\
        &       & 0.8 & 20.8 & 13.6 \\
        &       & 0.9 & 20.8 & 15.1 \\
        &       & 1.0 & 21.8 & 18.4 \\
\end{tabular}
\caption{The estimated ``bulk'' permittivity $\epsilon_b$ and
  zero-frequency  permittivity $\epsilon(\omega=0)$ values for the
  different cavity sizes $R$, reduced densities $\rho^*$, and
  $\kappa$.}
\label{T:er}
\end{table}

Results for the smaller cavity ($R=3$ nm), i.e. higher overall density
are shown in Fig.~\ref{F:ep30}. As is already clear from
Fig.~\ref{F:rhoQ30}, Eq.~(\ref{E:epr}) is no longer applicable, because the
strong oscillations in $Q_\text{ind}(r)$ go repeatedly through the value
$-e$ and a ``bulk''-like regime is never reached. Hence only the
results from the fluctuation formula (\ref{E:ep}) 
are shown. The behaviour as a function of $\kappa$ qualitatively
confirms one of the observations already made at the lower density
($R=4$ nm) shown in Fig.\ref{F:ep40}, namely that the ``bulk'' values
of $\epsilon$ agree within statistical errors for $\epsilon'>1$
($\kappa=0.8$, 0.9, and 1), but are roughly a factor 2 lower than the
value measured for $\epsilon'=1$. The error bars on the latter are much
larger than those associated with the data for the polarisable
embedding medium.

At the highest density ($R=2.5$ nm), $\epsilon(r)$ oscillates roughly
in phase with the density oscillations. A proper ``bulk'' regime is
never reached for the $N=1000$ particle system, but one can extract a
rough value of $\epsilon_b$ around which $\epsilon(r)$
oscillates. These values, given in Table~\ref{T:er}, depend only weakly
on $\kappa$, except for $\kappa=0.5$ (vacuum outside the cavity), when
$\epsilon_b$ is roughly a factor of two larger than for $\kappa>0.5$;
the very noisy permittivity profile for $\kappa=0.5$ is not shown in
Fig.~\ref{F:ep25}. The best estimates of $\epsilon_b$ as a function of
cavity radius $R$ and of $\kappa$ are listed in Table~\ref{T:er}. The
permittivity of the confined fluid is strongly reduced compared to its
value in a uniform (bulk) fluid at the same density and temperature.

\section{Relaxation}

\begin{figure}[thb]
\centering
\includegraphics[width=8cm]{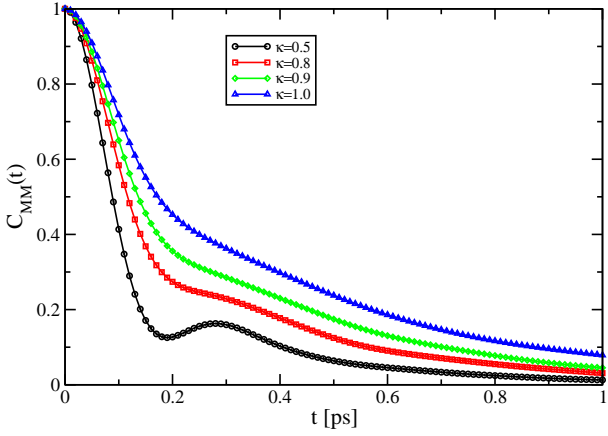}
\caption{The total dipole autocorrelation function $C_{MM}(t)$ versus time
  for $R=4$ nm, $\mu^*=2$, $0.5 \leq \kappa \leq 1$.}
\label{F:CMM40}
\end{figure}
\begin{figure}[thb]
\centering
\includegraphics[width=8cm]{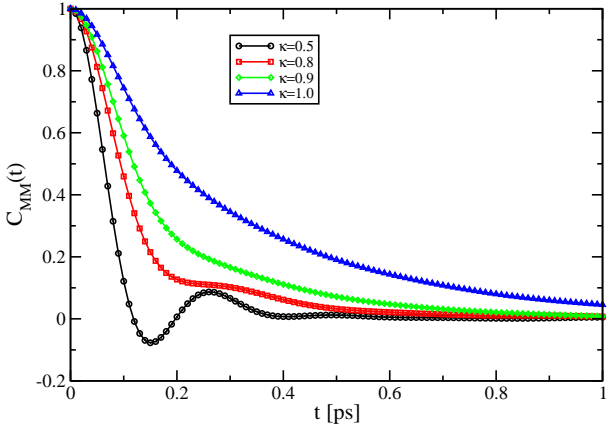}
\caption{The total dipole autocorrelation function $C_{MM}(t)$ versus time
  for $R=3$ nm, $\mu^*=2$, $0.5 \leq \kappa \leq 1$.}
\label{F:CMM30}
\end{figure}

The dielectric response of a polar sample is characterised by the
frequency-dependent complex dielectric permittivity $\epsilon(\omega) =
\epsilon_1(\omega) + \imath \epsilon_2(\omega)$. Within the linear
response regime the latter is determined by the Laplace transform of
the dynamical response function $\Phi_{MM}(t)$ which relates the
induced total dipole moment of the sample to a time-dependent external
field~\cite{Madden:1984ACP,Book:Hansen-McDonald3}: 
\begin{equation}
\langle \Delta \vec{M}(t) \rangle = \int_{-\infty}^t \Phi_{MM}(t - t')
\vec{E}_\text{ext}(t') d t'
\end{equation}
where $\Phi_{MM}$ is a scalar for a spherical sample. According to the
standard rules of linear response~\cite{Book:Hansen-McDonald3}: 
\begin{equation}
\label{E:PhiM}
\begin{split}
\Phi_{MM}(t) & = - \beta \langle \dot{\vec{M}}(t) \cdot
\vec{M}(0) \rangle \\
 & = - \beta \dot{C}_{MM}(t) \langle M^2\rangle
\end{split}
\end{equation}
where the dot denotes a time derivative and $C_{MM}(t)$ is the normalised total
dipole moment correlation function of the unperturbed sample:
\begin{equation}
\label{E:CMM}
C_{MM}(t) = \frac{ \langle \vec{M}(t) \cdot \vec{M}(0)
  \rangle}{\langle M^2\rangle} 
\end{equation}
The complex susceptibility is:
\begin{equation}
\tilde{\chi}_{MM}(z) = \int_0^\infty \Phi_{MM}(t) e^{\imath z t} d t
\end{equation}
\begin{figure}[thb]
\centering
\includegraphics[width=8cm]{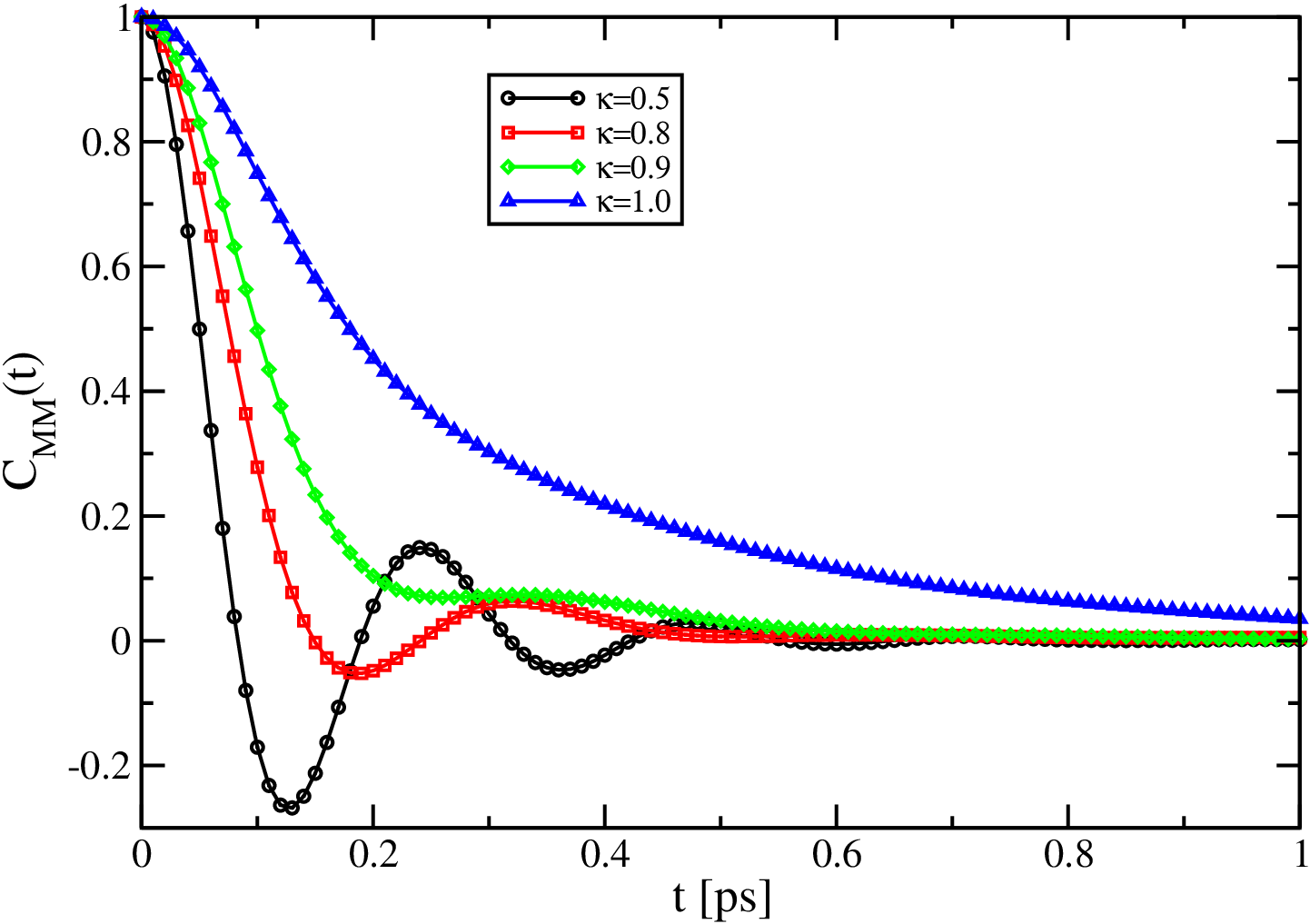}
\caption{The total dipole autocorrelation function $C_{MM}(t)$ versus time
  for $R=2.5$ nm, $\mu^*=2$, $0.5 \leq \kappa \leq 1$.}
\label{F:CMM25}
\end{figure}
where $z=\omega + \imath \varepsilon$, and:
\begin{equation}
\label{E:chi}
\lim_{\varepsilon \rightarrow 0^+} \tilde{\chi}_{MM}(z) = \chi_1(\omega) +
\imath \chi_2(\omega)
\end{equation}
According to the fluctuation-dissipation theorem, a direct consequence
of Eq.~(\ref{E:PhiM}):
\begin{equation}
\tilde{\chi}_{MM}(z) = \beta \langle M^2\rangle \left[
  C_{MM}(t=0) + \imath z \tilde{C}_{MM}(z) \right]
\end{equation}
This implies that the spectrum of the correlation function $
C_{MM}(t)$ is related to the imaginary part of the susceptibility
\begin{equation}
\begin{split}
\hat{C}_{MM}(\omega) & = \frac{1}{2 \pi} \int_{-\infty}^\infty e^{\imath
  \omega t}  C_{MM}(t) d t \\
& = \frac{k_B T}{\pi \omega} \frac{\chi_2(\omega)}{\langle M^2\rangle}
\end{split}
\end{equation}
Finally, for the system under consideration, i.e. a polar fluid
confined to a spherical cavity surrounded by a dielectric continuum of
permittivity $\epsilon'$, the frequency-dependent permittivity of the
sample is related to the complex susceptibility
by~\cite{Titulaer:1974JCP} : 
\begin{equation}
\label{E:epw}
\begin{split}
\frac{\epsilon(\omega)-1}{\epsilon -1} \frac{2 \epsilon' + \epsilon}{2
  \epsilon' + \epsilon(\omega)} 
& = \frac{k_B T}{\langle M^2\rangle} \tilde{\chi}_{MM}(\omega) \\
& = \frac{k_B T}{\langle M^2\rangle} \left[ \chi_1(\omega) + \imath
  \chi_2(\omega) \right] 
\end{split}
\end{equation}
where  $\epsilon \equiv \epsilon(\omega=0)$ is the static
permittivity of the sample.

The static permittivity $\epsilon$ is given by the $\omega \rightarrow
0$ limit of Eq. (\ref{E:epw}) which results in:
\begin{equation}
\frac{(\epsilon-1)(2 \epsilon' + 1)}{2 \epsilon' + \epsilon} = 
\frac{4 \pi \beta}{3} \frac{\langle M^2 \rangle}{V}
\end{equation}
\begin{figure}[thb]
\centering
\includegraphics[width=8cm]{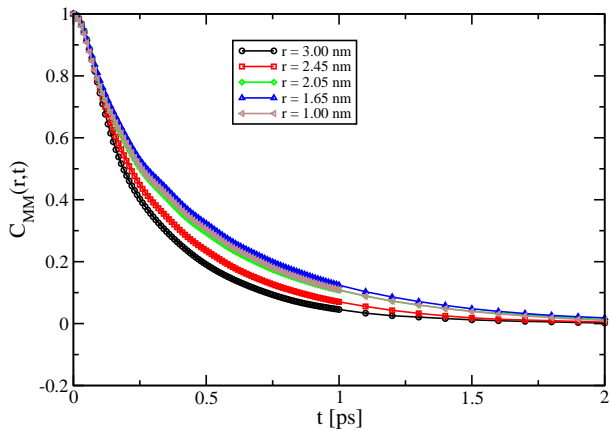}
\caption{The dipole autocorrelation function $C_{MM}(r,t)$ versus time
  for $R=3$ nm, $\mu^*=2$, $\kappa =1$ and several values of $r$.}
\label{F:CMMr10}
\end{figure}
Thus $\epsilon(\omega=0)$ is determined by the fluctuation of the
total dipole moment $\vec{M}$ of the spherical sample. The result
differs from the ``bulk'' value $\epsilon_b$ determined as explained
in Sect.~\ref{S:static}, from the profile $\epsilon(r)$ calculated
from Eq.~(\ref{E:ep}). It characterises the global response of the
polar fluid trapped in the cavity rather than the local response, away
from the confining surface. The values of $\epsilon$ are seen to lie
systematically below those of $\epsilon_b$, particularly so in the
case $\kappa=0.5$.

Hence all information required to compute
$\epsilon(\omega)$ is contained in the total dipole autocorrelation
function (\ref{E:CMM}). MD results for $C_{MM}(t)$ are shown in
Figures \ref{F:CMM40} -- \ref{F:CMM25} for the three cavity radii
$R=4$, 3, and 2.5 nm, and four values of
$\kappa$.
The correlation
functions are seen to relax to zero over a time scale of about 1 ps,
which is an order of magnitude shorter than the relaxation time
observed in the bulk for the same model~\cite{Ballenegger:2004MP}. The
most striking feature is the strong sensitivity of $C_{MM}(t)$ to the
polarisability of the confining medium, i.e. to $\kappa$. For all three
radii, the relaxation is slowest for $\kappa=1$ (metallic boundary),
and becomes faster as $\kappa$ decreases. Marked oscillations appear
when $\kappa=0.5$ ($\epsilon'=1$) particularly so at the highest
density ($R=2.5$ nm). Simulations carried out on smaller samples of
$N=250$ dipoles show that the relaxation patterns appear to be
independent of the sample size characterised by $R$ and $N$, provided
the reduced overall density $\rho^*=\rho \sigma^3$ is the same. The
decrease of the relaxation time of $C_{MM}(t)$ with $\kappa$ agrees
with the behaviour predicted for a Debye
dielectric~\cite{Neumann:1983CPL}. There is no obvious explanation for
the oscillation observed for $\kappa=0.5$. These oscillations are
indicative of collective behaviour reminiscent of the dipolaron mode
observed in MD simulations of longitudinal dipolar fluctuations at
finite wavenumber in bulk model polar fluids~\cite{Pollock:1981PRL}.In
an effort to gain a 
better understanding of the dipolar relaxation, we have also computed
the normalised correlation functions:
\begin{figure}[thb]
\centering
\includegraphics[width=8cm]{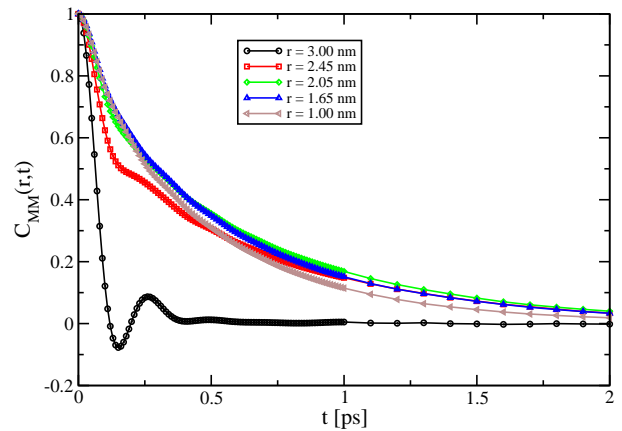}
\caption{The dipole autocorrelation function $C_{MM}(r,t)$ versus time
  for $R=3$ nm, $\mu^*=2$, $\kappa =0.5$ and several values of $r$.}
\label{F:CMMr05}
\end{figure}
\begin{equation}
C_{MM}(r,t) = \frac{ \langle \vec{M}(r,t) \cdot \vec{M}(r,0)
  \rangle}{\langle |\vec{M}(r)|^2\rangle} 
\end{equation}
where $\vec{M}(r,t)$ is the instantaneous total dipole moment of all
the molecules contained inside a sphere of radius $r \leq R$. The MD
data are shown in Figures \ref{F:CMMr10} and \ref{F:CMMr05} for
$\kappa=1$ and $\kappa=0.5$ respectively. In the former (metallic
boundary) case $C_{MM}(r,t)$ changes only moderately with $r$. This is
not totally unexpected, since the ``bulk'' permittivity of the sample
is relatively large under these conditions ($\epsilon \simeq 12$), as
seen from Fig.~\ref{F:ep30}, and hence the dielectric discontinuity
is not too strong relative to the metallic embedding medium. The
situation is very different when $\kappa=0.5$ (Fig.~\ref{F:CMMr05}). In
this case $C_{MM}(r,t)$ changes relatively little with $r$, and decays
monotonically except when $r=R$ (corresponding to the autocorrelation
function of the total dipole moment of the sample), when $C_{MM}$
decays much faster and oscillates. Thus the dynamics of the molecular
dipole moments inside the outer shell, in direct contact with the
surface separating the confined fluid from vacuum, has a dramatic
effect on the total dipole correlation function. 

\begin{figure}[thb]
\centering
\includegraphics[width=8cm]{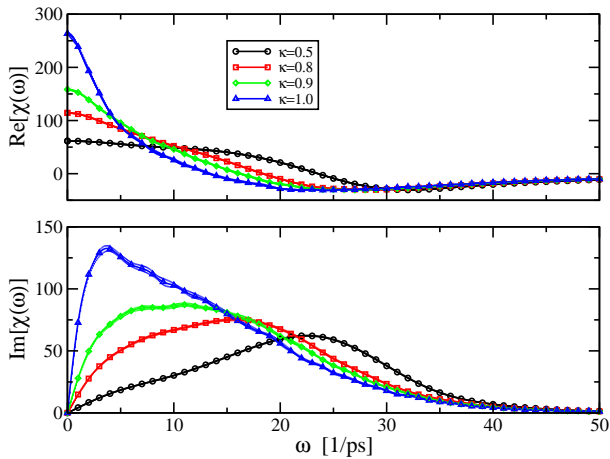}
\caption{The real and imaginary part of the complex susceptibility as
  a function of the frequency $\omega$ ($R=3$nm).}
\label{F:chiw30}
\end{figure}

Real and imaginary parts of the complex susceptibility (\ref{E:chi})
are plotted in Fig.~\ref{F:chiw30} for the cavity of radius $R=3$ nm,
and 4 values of $\kappa$. $\chi_1(\omega)$ and $\chi_2(\omega)$ vary
with $\omega$ in a manner reminiscent of bulk
behaviour~\cite{Pollock:1981PRL}. However they are rather sensitive to
$\kappa$, i.e. to the permittivity of the surrounding medium. On the
contrary the real and imaginary parts of the frequency-dependent
permittivity defined by Eq.~(\ref{E:epw}), are remarkably insensitive
to $\kappa$, except for $\epsilon_1(\omega)$ in the static ($\omega
\rightarrow 0$) limit, as illustrated in Fig.~\ref{F:epw30}. This
insensitivity to $\kappa$ reflects the fact that, contrary to
$\tilde{\chi}_{MM}(\omega)$, $\epsilon(\omega)$ measures the response
of the polar fluid to the local, internal field.

The resulting Cole--Cole plots, shown in Fig.~\ref{F:Cole30}, differ
strongly from the semi-circular shape of the simple Debye theory, as
one might expect. The high-frequency part is very insensitive to
$\kappa$, while the differences in the low-frequency range reflect the
significant differences in static values of $\epsilon$.

\section{Conclusion}
We have reported the first systematic attempt to investigate  the
dependence of the structure, static and dynamic correlations and
response of a drop of polar fluid confined to a spherical cavity, on
the permittivity $\epsilon'$ of the embedding medium. The work extends
earlier investigations which were restricted to the case of a
non-polarisable external medium
($\epsilon'=1$)~\cite{Senapati:1999JCP,Ballenegger:2005JCP} by
treating the interactions between the charge distribution associated
with the extended dipoles of the trapped fluid, and the image charges
in an approximate, but accurate way, which provides an efficient
alternative to the more cumbersome variational method proposed
elsewhere~\cite{Allen:2001PCCP}. The present treatment of
electrostatic boundary conditions for the electric field of individual
charges within the confined sample does not rely on macroscopic
reaction field considerations, but is restricted to the spherical
geometry. 

The MD simulations were run for samples of $N=1000$ polar molecules
confined to cavities of radii $R=4$, 3, and 2.5 nm, which amount to
effective densities of $\rho^*=0.23$, 0.53, and 0.92. Some test runs
were carried 
out for a smaller sample of $N=250$ molecules, and no significant
$N$-dependence was observed. The larger $N=1000$ particle system
allows a ``bulk'' regime to be reached within a substantial fraction
of the accessible volume (say up to $r \simeq R/2$), except for the
smallest cavity (i.e. highest effective density $\rho^*=0.92$). All
calculations were made with a reduced extended dipole moment
$\mu^*=2$, comparable to that of water.

\begin{figure}[thb]
\centering
\includegraphics[width=8cm]{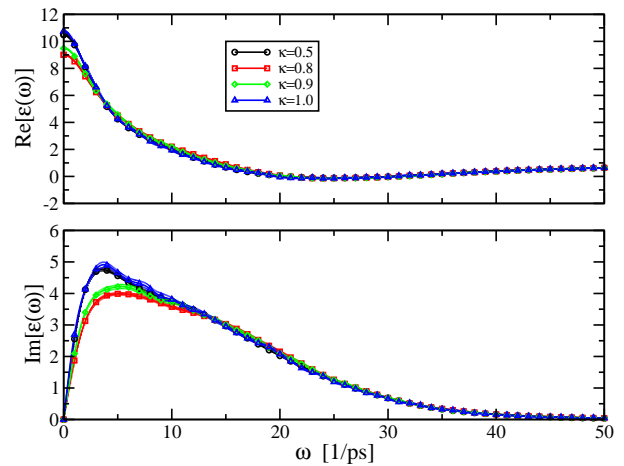}
\caption{The real and imaginary part of the complex permittivity as
  a function of the frequency $\omega$ ($R=3$nm).}
\label{F:epw30}
\end{figure}
\begin{figure}[thb]
\centering
\includegraphics[width=8cm]{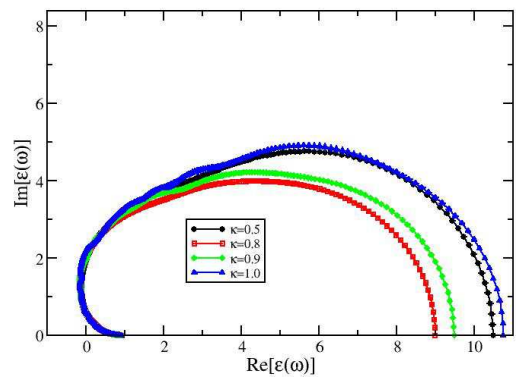}
\caption{Cole--Cole plot: imaginary versus real part of the
  permittivity ($R=3$nm).}
\label{F:Cole30}
\end{figure}

The main conclusions to be drawn from our  MD data may be summarised
as follows:
\begin{enumerate}
\item[{\em a)}]
The structural properties embodied in the density profiles $\rho(r)$
and the order parameter profiles $\langle P_2(\hat{\mu} \cdot
\hat{r})\rangle$ are remarkably insensitive to the embedding medium,
i.e. to $\kappa$. The density profiles show significantly less
structure than their $\mu=0$ counterparts. The order parameter profiles
point to a preferential alignment of the dipoles parallel to the
confining surface, as already reported in earlier
studies of related systems~\cite{Senapati:1999JCP,Zhang:1995JPC}.
\item[{\em b)}]
The static permittivity profiles $\epsilon(r)$ may be calculated from
the generalised Kirkwood fluctuation relation
(\ref{E:ep})\cite{Ballenegger:2005JCP}, or by measuring the
polarisation profile $P(r)$, or charge profile $Q(r)$ induced by an
``external'' charge placed at the centre of the spherical cavity
(cf. Eq.~(\ref{E:epr})). The latter method can only be implemented at
the lowest density ($R=4$ nm) and points to a significant
non-linearity compared to the predictions of the fluctuation formula,
when the ``external'' charge is the proton charge. The fluctuation
formula yields oscillatory $\epsilon(r)$ profiles which essentially
reflect the oscillations in the density profiles $\rho(r)$. At the two
lower densities ($R=4$ and 3 nm) the oscillations are sufficiently
damped away from the confining surface for a ``bulk'' regime to be
reached inside the cavity allowing the definition of a ``bulk''
permittivity $\epsilon_b$. The latter turns out to be relatively
insensitive to $\kappa$, except for $\kappa=0.5$ (cavity surrounded by
vacuum), which leads to substantially larger values of
$\epsilon_b$. At the highest density ($R=2.5$ nm), the oscillations
of $\epsilon(r)$ extend up to the centre , so that no proper ``bulk'' regime
is reached . A rough estimate of $\epsilon_b$ may be extracted by
averaging over  oscillations. In all cases the ``bulk'' permittivity
inside the cavity is strongly reduced relative to the values expected
for a uniform (genuinely bulk) fluid under comparable conditions, an
observation also made in earlier
work~\cite{Senapati:1999JCP,Zhang:1995JPC}. 
\item[{\em c)}]
While the static properties of a confined drop of polar fluid turn out
the be surprisingly insensitive to the permittivity of the external
medium except when $\epsilon' \rightarrow 1$, the dynamical
properties depend much more on $\epsilon'$ (or equivalently
$\kappa$). The most striking illustration is the correlation function
$C_{MM}(t)$ of the total dipole moment of the sample, which relaxes
faster when $\kappa \simeq 0.5$ (cf. Figs.~\ref{F:CMM40} --
\ref{F:CMM25}). As also noted in earlier work~\cite{Zhang:1995JPC} on
confined water, the relaxation of $C_{MM}(t)$ is much faster than under
comparable bulk conditions~\cite{Ballenegger:2004MP}. While such a
behaviour may be rationalised by the absence of long-range dipolar
interactions with distant molecules in the case of the confined
system, which may lead to a lower collective ``inertia'' compared to
the bulk, a detailed theoretical interpretation of the observation is
still lacking.
\item[{\em d)}]
The complex, dynamical permittivity was estimated from
Eq.~(\ref{E:epw}) The corresponding Cole--Cole plots differ
considerably from the semi-circular shape associated with exponential
Debye relaxation, as was to be expected from the complex relaxation
pattern of the $C_{MM}(t)$ correlation function. While the high
frequency regions of the real and imaginary parts of the permittivity
are quite insensitive to the value of $\kappa$, significant deviations
occur in the low-frequency regime (cf.~Figs.~\ref{F:chiw30} and
\ref{F:epw30}) which are reflected in the right-hand parts of the
Cole--Cole plots in Fig.~\ref{F:Cole30}. The zero-frequency limits
$\epsilon=\epsilon(\omega=0)$ of the dynamical permittivity differ
substantially from the ``bulk'' limits $\epsilon_b$ of the static
permittivity profiles $\epsilon(r)$, as shown in Table~\ref{T:er},
which is not surprising since $\epsilon$ and  $\epsilon_b$ measure
``global'' and ``local'' responses respectively.
\end{enumerate}
A theoretical analysis of the dynamical properties of confined polar
fluids is left for future work. We also plan to explore the static and
dynamic properties of polar fluids in narrow pores (one-dimensional
confinement) and in slits (two-dimensional confinement).

\begin{acknowledgments}
R.B. acknowledges the support of EPSRC within the Portfolio Grant
Nr. RG37352. We thank Vincent Ballenegger for his interest and advice.
\end{acknowledgments}

\appendix

\section{}
In this appendix we derive the electrostatic potential inside
a spherical cavity of radius $R$ and permittivity $\epsilon$, which
results from the polarisation of an infinite medium with permittivity
$\epsilon'$ surrounding the cavity due to an internal charge
distribution. 

The electrostatic potential at $\vec{r}$, due to a single charge $Q$ at
position $\vec{d}$ inside the cavity, can be formally expanded in
Legendre Polynomials $P_l$ as:
\begin{equation}
\begin{split}
\Phi^\text{in} & = \frac{Q}{4 \pi \epsilon} \frac{1}{|\vec{r} - \vec{d}|} + 
\sum_{l=0}^\infty A_l r^l P_l(\cos \theta) ;~~~ r<R\\
\Phi^\text{out} & = \sum_{l=0}^\infty \frac{B_l}{r^{l+1}} P_l(\cos \theta); ~~~~~~~~~~~~~~~~~~~~~   r > R
\end{split}
\end{equation}
where $\cos \theta = \hat{r} \cdot \hat {d}$, and the expansion
coefficients $A_l$ and $B_l$ follow from the boundary conditions of a
continuous tangential and a discontinuous normal electric field at the
border of the cavity $r=R$:
\begin{equation}
\begin{split}
 \frac{\partial \Phi^\text{in}}{\partial \theta}  & = 
  \frac{\partial \Phi^\text{out}}{\partial \theta}   \\ 
 \epsilon  \frac{\partial \Phi^\text{in}}{\partial r} & = 
  \epsilon' \frac{\partial \Phi^\text{out}}{\partial r}  
\end{split}
\end{equation} 
By expanding the direct term $1/|\vec{r} - \vec{d}|$ of the internal
electrostatic field $\Phi^\text{in}$ at the cavity wall in Legendre
polynomials, solving these boundary conditions is straightforward and
one finds: 
\begin{equation}
\begin{split}
\Phi^\text{in} & = \frac{Q}{4 \pi \epsilon} \left[ \frac{1}{|\vec{r} -
    \vec{d}|} + (1 - 2 \kappa) \sum_l \frac{l+1}{l+\kappa}
  \frac{d^l r^l}{R^{2 l + 1}} P_l(\cos \theta) \right]\\
\Phi^\text{out}& =  \frac{Q \kappa}{4 \pi \epsilon'} 
\sum_l \frac{2 l+1}{l+\kappa} \frac{d^l}{r^{l + 1}} P_l(\cos \theta)
\end{split}
\end{equation}
where we introduced $\kappa \equiv \epsilon'/(\epsilon+\epsilon')$.
This result can be simplified by rewriting the fraction appearing
inside the summations:
\begin{equation}
\begin{split}
\Phi^\text{in} = & \frac{Q}{4 \pi \epsilon} \left[ \frac{1}{|\vec{r} -
    \vec{d}|} + (1 - 2 \kappa) \sum_l \frac{d^l r^l}{R^{2 l + 1}}
    P_l(\cos \theta) + \right. \\ 
& \left. \frac{(1 - 2 \kappa)(1-\kappa)}{\kappa R}
    \sum_l \frac{\kappa}{l+\kappa}  
  \frac{d^l r^l}{R^{2 l}} P_l(\cos \theta) \right]
\end{split}
\end{equation}
\begin{equation}
\begin{split}
\Phi^\text{out} = & \frac{Q \kappa}{4 \pi \epsilon'} \left[
2 \sum_l  \frac{d^l}{r^{l + 1}} P_l(\cos \theta) + 
\right. \\ 
& \left. \frac{1 - 2 \kappa}{\kappa r} \sum_l \frac{\kappa}{l+\kappa}
\frac{d^l}{r^{l}} P_l(\cos \theta) \right]
\end{split}
\end{equation}

In both cases the first summation represents the expansion in Legendre
polynomials of an inverse distance. In the latter this is the distance
$|\vec{r} - \vec{d}|$ to the original charge. In the former, however,
this is the distance to a location $\vec{D} \equiv =(R/d)^2 \vec{d}$
outside the cavity. The second summation can be simplified by
expanding the Legendre polynomial in terms of $\exp(\imath n \theta)$
(See Ref.~\cite{Book1:Gradshteyn-Ryzhik}, Eq.[8.911.4]). This results
in:
\begin{equation}
\begin{split}
\Phi^\text{in} = & \frac{1}{4 \pi \epsilon} \left[ \frac{Q}{|\vec{r} -
 \vec{d}|} + (1 - 2 \kappa) \frac{Q (R/d)}{|\vec{r} -
    \vec{D}|} \right] + \\
 & \frac{Q(1-2\kappa)}{4 \pi \epsilon' R} F_1(\kappa,\frac{1}{2},\frac{1}{2},1+\kappa;x
 e^{\imath \theta},x e^{-\imath \theta})
\end{split}
\end{equation}
\begin{equation}
\begin{split}
\Phi^\text{out} = &\frac{1}{4 \pi \epsilon'} \frac{2 \kappa Q}{|\vec{r} -
 \vec{d}|} + \\
 & \frac{Q(1-2\kappa)}{4 \pi \epsilon' r}
 F_1(\kappa,\frac{1}{2},\frac{1}{2},1+\kappa;y e^{\imath \theta},y
 e^{-\imath \theta})  
\end{split}
\end{equation}
where we introduced $x \equiv r/D$ and  $y \equiv d/r$, while $F_1$ is
a hyper-geometric function in two variables (See
Ref.~\cite{Book1:Gradshteyn-Ryzhik}, Eq.[9.180.1]). Note that the
location $\vec{D}$ corresponds to the position where a single image
charge should be placed in the case of metallic boundary
conditions~\cite{Book:Jackson}.

Although one can in principle calculate or tabulate the
hyper-geometric function, one can show that in the present case of a
confined dipolar fluid this term can safely be neglected. In doing so,
we approximate the induced electrostatic potential field by a single
external image charge. Note that this approximation is exact in the
case of a vacuum outside ($\kappa=1/2$) and in the case of metallic
boundary conditions ($\epsilon' \rightarrow \infty$). 

\begin{figure}[thb]
\centering
\includegraphics[width=8cm]{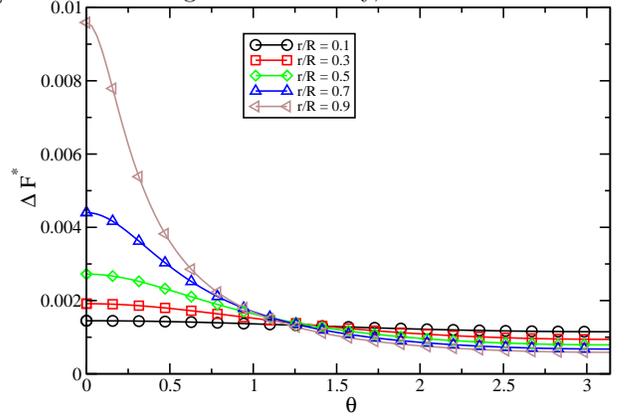}
\caption{The absolute error in the reduced force $\Delta F^*$ of the
  force of a unit charge at a distance $d/R=0.9$ from the origin on a
  unit charge at a position $(r,\theta)$ inside a cavity with radius
  $R=2.5$ nm and $\kappa=0.8$.}
\label{F:App}
\end{figure}

There are two intuitive arguments for making this
approximation. Firstly, the field, except near the cavity wall, is 
mainly determined by the local charge distribution inside the cavity,
rather than the image charge distribution arising from the
polarisation. The second reason is that, since we consider a fluid of
extended dipoles where the charge separation is roughly a third of the
particle diameter, the two neglected parts of the induced charge
distribution of both charges that form a dipole will cancel to a large
extent. 

In order to illustrate that the error made by the approximation is indeed
small, we place a unit charge at a distance $d/R=0.9$ from the origin
in a cavity of radius $R=2.5$ nm and $\kappa=0.8$, and measure the
reduced force $F^*$ in terms of the force between two unit charges at
a distance $\sigma$. A second charge is put at various distances $r/R$
from the origin of the cavity, the result of which is shown in
Fig.~\ref{F:App}. The error increases both with decreasing cavity radius,
and on approaching the cavity wall with one or both charges. Note, that
although the absolute error increases also when the charges get closer, the
relative error in that case decreases due to the singular behaviour of the
direct interaction between the charges.

\end{document}